\documentstyle[epsf]{article}

 1
 1
 1

\title{Radiative Corrections to $W$ Pair Production
       at High Energies%
\thanks{Presented by Thomas Mannel}
}
\author{{\sc H. Anlauf, A. Himmler, P. Manakos, T. Mannel} \\
        Institut f\"ur Kernphysik, 
        Schlossgartenstr.~9, D--64289 Darmstadt, 
        Germany, \vspace*{1mm} \\
        {\sc H. Dahmen} \\
        Universit\"at Siegen, 
        Adolf Reichwein Stra\ss{}e, D--57076 Siegen, 
        Germany}
\begin{document}
\maketitle

\hbox{\vbox to0pt{\vss\hbox to\hsize{\hfill\large IKDA 93/25}
      \vskip55mm }}
\vspace*{-11mm}

\thispagestyle{empty}
\unitlength1mm

\section{Introduction}
The spectacular success of the high energy electron positron colliders
SLC at SLAC and LEP at CERN has confirmed the predictions of the
Standard Model (SM) for the interactions between the gauge bosons and
the fermions event at the level of electroweak radiative corrections.
However, the non-abelian structure of the gauge sector of the SM with
its couplings between the electroweak gauge bosons have not been
tested directly. In addition, the origin of electroweak symmtry
breaking giving longitudinal components to the electroweak gauge
bosons, is still obscure.

Anomalous couplings will disturb the extensive gauge cancellations
taking place in the standard model, and possible new physics will show
up in particular in the cross section for the production of
longitudinally polarized $W$ bosons. On the other hand, possible new
physics is already severely constrained by LEP100 data, and the
effects to be expected at a 500 GeV linear collider are small.

In order to extract the small effects of physics beyond the SM one
has to have a precise knowledge of the radiative corrections within
the SM.
The electroweak radiative corrections to the production of on-shell
$W$'s to one-loop order are by now well established \cite{RC}.
The influence of the finite width of the $W$'s has been investigated
in~\cite{Finite-Width}.  Also, the higher order QED corrections have
been calculated in the leading log approximation (LLA)~\cite{CDMN91}.

However, the experimental reconstruction of the $W$'s is complicated
by the fact that they may decay either into leptons with an escaping
neutrino, or into hadrons, where the jet energies may be poorly known
due to undetected particles.  In addition, the radiative corrections
due to emission of photons produce a systematic shift of the effective
center of mass energy towards smaller values.  Such effects may best
be studied with the help of a Monte Carlo event generator.

In the present contribution we present first results obtained with
Monte Carlo event generator {\tt WOPPER} which allows to simulate $W$
pair production including radiative corrections and effects from
finite $W$ width. We shall consider two applications, the
reconstruction of the $W$ boson helicities from the semileptonic final
states and the effects of the radiative corrections and the finite
width of the $W$ bosons on the total cross section.

\section{The Monte Carlo {\tt WOPPER} }
The Monte Carlo Event generator {\tt WOPPER} is capable of a full
simulation of the cross section for $e^+ e^- \to 4$ fermions +
$n\gamma$ via the resonant channel containg two $W$ bosons. The finite
width of the $W$ bosons is included as well as QED radiative
corrections in all orders of the leading logarithmic approximation
(LLA). At very high energies these corrections are indeed the ones
which are numerically most important, since
\begin{equation}
  \frac{\alpha}{\pi} \log \left(\frac{s}{m_e^2}\right)  \approx 6\%
  \qquad  \mbox{(at LEP200 and EE500 energies)}
\end{equation}

The LLA is conveniently incorporated using the so called structure
function formalism \cite{Structure}. In this formalism the expression
for the radiatively corrected cross section reads
\begin{equation}
  \sigma(s) = \int\limits_0^1 dx_+ dx_- \;
  D(x_+,Q^2) D(x_-,Q^2) \; \hat\sigma(x_+ x_- s) \; ,
  \label{eq:factorization}
\end{equation}
where $\hat\sigma$ is the Born level cross section of the hard process,
$D(x,Q^2)$ are the structure functions for initial state radiation, and
$Q^2 \sim s$ is the factorization scale. The structure function satisfies the 
evolution equation
\begin{equation}
\label{eq:DGLAP}
   Q^2 \frac{\partial}{\partial Q^2} D(x,Q^2)
      =  \frac{\alpha}{2\pi}
             \int\limits_x^1 \frac{dz}{z} \left[P_{ee}(z)\right]_+
                 D\left(\frac{x}{z},Q^2\right) 
\quad \mbox{ with } \quad
  P_{ee}(z)  =  \frac{1+z^2}{1-z}
\end{equation}
with the initial condition $ D(x,m_e^2) = \delta(1-x) $.  The solution
to eq.~(\ref{eq:DGLAP}) automatically includes the exponentiation of
the soft photon contributions as well as a resummation of the large
logarithms of the form $\ln (s/m_e^2)$ from multiple hard photon
emission.

The radiatively corrected cross section (\ref{eq:factorization}) is
implemented in a Monte Carlo event generator by solving 
(\ref{eq:DGLAP}) by iteration. 
This procedure is well known from the corresponding QCD applications
\cite{QCD} and, as
a by-product of the algorithm used, the four-momenta of the
radiated photons may be generated explicitly. For more details we refer
the reader to \cite{Good-Stuff}.

The Monte Carlo {\tt WOPPER} also includes the effects of the finite
width of the $W$ bosons. To introduce finite width for the decaying
$W$ bosons one has various possibilities \cite{Width}. The one chosen 
in the Monte Carlo {\tt WOPPER} is to start from an off-shell cross 
section $\sigma_{os}$ for the process $e^+ e^- \to W^+ W^-$ where the 
momenta $k_\pm$ of the $W$ bosons are not on shell
\begin{equation} 
\sigma_{os} = \sigma_{os} (s; k_+^2, k_-^2) \, 
\end{equation}
which is obtained by continuating the momenta of the $W$ to off-shell
values. The cross section for 
the process $e^+ e^- \to 4$ fermions is then obtained by convoluting 
$\sigma_{os}$
with propagators for the $W$ bosons multiplied by the decay probability 
for the subsequent $W$ decay
\begin{equation}
  \sigma = 
  \int \frac{ds_+}{\pi} \frac{ds_-}{\pi} \;
  \frac{\sqrt{s_+} \,\Gamma_W(s_+)}
       {(s_+ - M_W^2)^2 + s_+ \Gamma_W^2(s_+)}
  \frac{\sqrt{s_-} \,\Gamma_W(s_-)}
       {(s_- - M_W^2)^2 + s_- \Gamma_W^2(s_-)}
  \sigma_{os}(s; s_+, s_-)
  \label{eq:resonance-formula}
\end{equation}
where $\Gamma_W (s)$ denotes the width of the $W$ boson taken also off-shell. 
In the Monte Carlo {\tt WOPPER} the four fermion final states are generated 
according
to the distribution (\ref{eq:resonance-formula}). 
For more details see \cite{Good-Stuff}.

\section{Applications}

\subsection{Reconstruction of the $W$ Helicities}
The longitudinal modes of the electroweak gauge bosons play a specific
role in investigating the origin of electroweak symmetry breaking and
in extracting effects of physics beyond the SM. Hence a reconstruction
of the helicities of the $W$ bosons from their decay products is
mandatory.  There are three possibilities. Firstly, there may be a
hadronic final state which is not well suited for a reconstruction of
the $W$ helicities, since it requires to measure jet charge and/or jet
flavor, which is not yet feasible. The second possibility is a
leptonic final state which contains two neutrinos. Aside from the fact
that this channel is suppressed by a relatively small branching
fraction, the reconstruction is difficult due to initial state
radiation.  Finally, there is the possibility of a semileptonic final
state which is the best way of reconstructing $W$ helicities, since
the charge of the electron or muon may be determined and only one
neutrino is present which hence may be reconstructed.

The lowest order results for the reconstruction of the $W$ helicities
have been investigated in some detail in the past \cite{EE500}.
In order to reconstruct the neutrino momentum in the case where photons
are radiated we shall employ the fact that most of the radiated photons
are collinear with the beam. Due to this, most of the radiated
photons will be lost in the beam pipe and thus have to be treated
inclusively. Hence most of the photons have small transverse momentum 
and thus one may neglect their transverse momentum.
In the narrow width approximation one may solve the equations
\begin{equation}\label{eq:mom-cons}
p^x_\perp + p^l_\perp + p^{\nu}_\perp = 0 \; , \quad
(p^l + p^{\nu})^2 = M_W^2
\end{equation}
for the longitudinal momentum of the neutrino up to a twofold
ambiguity.  Here $p_l$ is the momentum of the charged lepton and $p_x$
is the total hadronic momentum.  The ambiguity in (\ref{eq:mom-cons})
is resolved by choosing the solution for which the invariant mass
$M_{WW}^2 = (p^{\nu} + p^x + p^l)^2 $ is closer to $\sqrt{s}$.

After having solved for the neutrino momentum we reconstruct the $W$
boson decay angle $\theta^*$ by a boost to the rest frame.  The cross
section corresponding to a definite helicity of the decaying $W^-$ is
obtained by convoluting the decay angular distribution with the
functions
\begin{equation}
f_{\pm} = {1 \over 2}(5 \cos^2{\theta^{\star}}
             -1 \mp 2\cos{\theta^{\star}})
 \; , \qquad
f_0 = 2-5\cos^2{\theta^{\star}}
\end{equation}
In fig.1 we compare the cross sections for the three helicities.  The
left column is the Born cross section, the radiatively corrected cross
section is the right column.  In both cases the above reconstruction
of the neutrino momentum is used.  This simulation is performed with
65 000 events in this channel which is an optimistic view of what may
be expected at EE500.  As may be seen from the figure the well known
enhancement of the longitudinal polarization due to radiative
corrections is clearly visible even at this level of statistics.

\begin{figure}[htb]
\begin{picture}(160,80)
\put(-10,0){
\makebox(80,80){\epsfxsize=8cm
    \leavevmode
    \epsffile{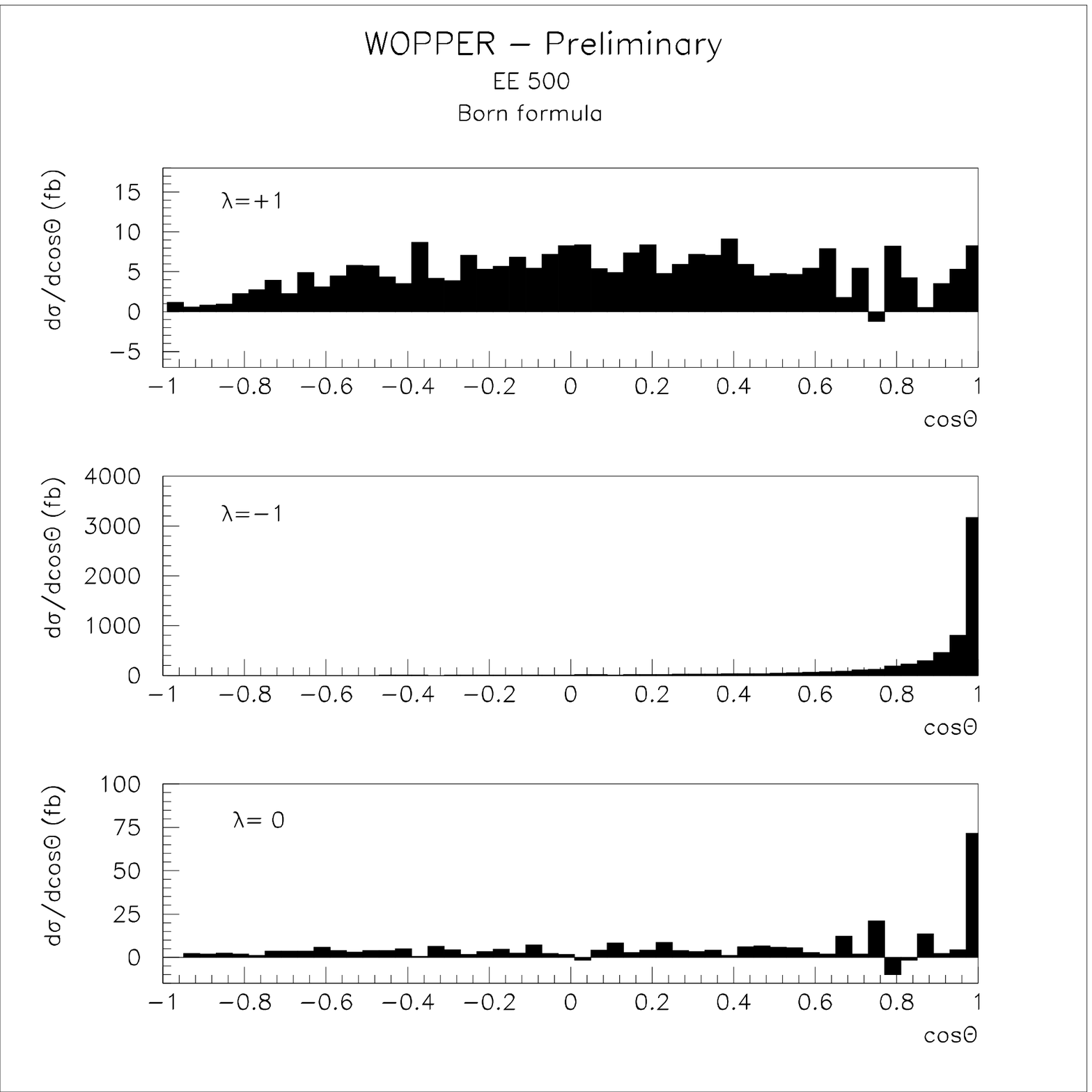}
}}
\put(75,0){
\makebox(80,80){\epsfxsize=8cm
    \leavevmode
    \epsffile{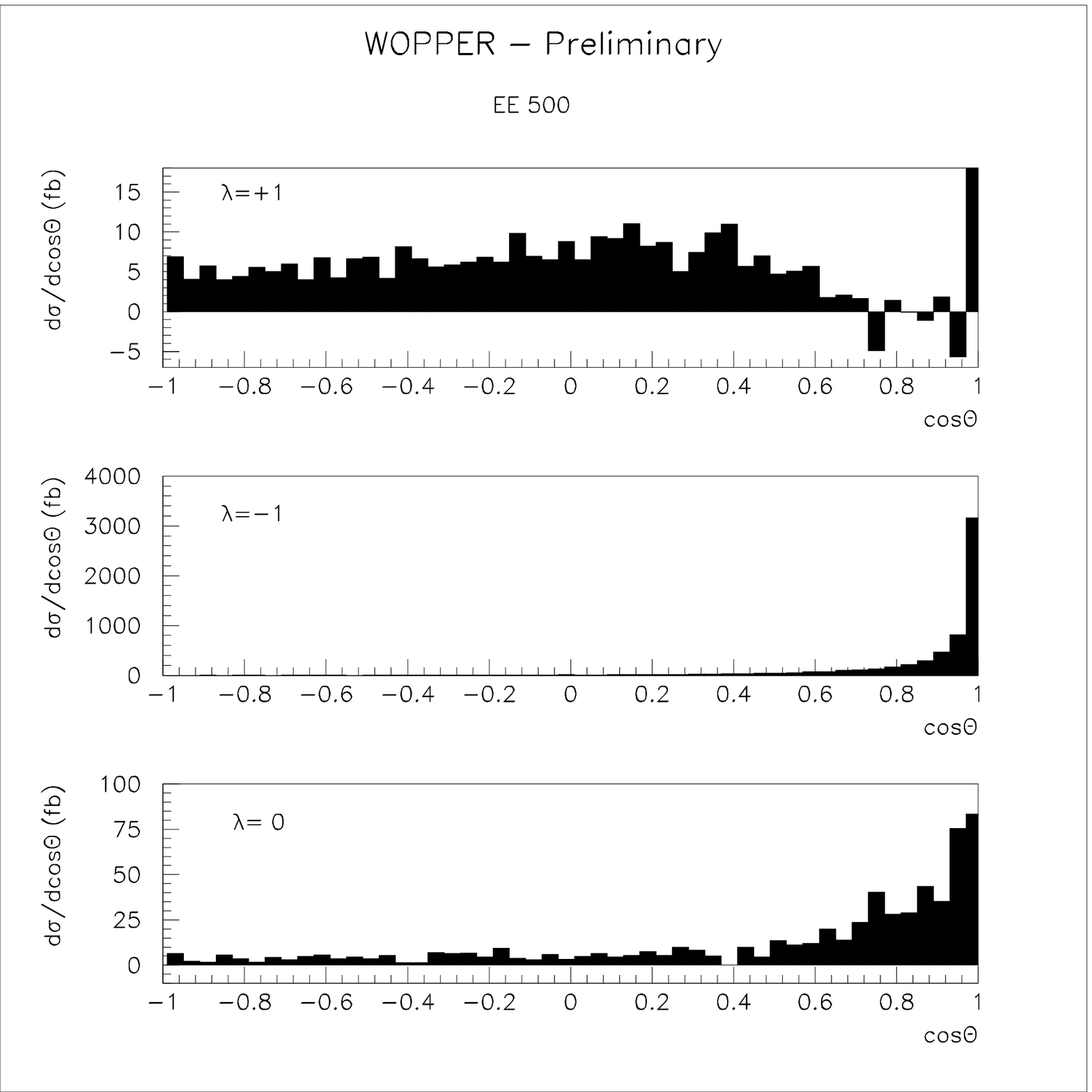}
}}
\end{picture}
\caption{Differential cross section for the $W$ helicities
       reconstructed as described in the text.
       Left column: Born cross section,
       Right column: cross section including QED corrections
       in LLA.}
\end{figure}

\subsection{Finite $W$ Width Effects}
Finally we shall discuss the effects of a finite $W$ width. In
principle we may study these effects for any exclusive quantity, since
{\tt WOPPER} is a full Monte Carlo generator. However, for the sake of
comparison with other work we will consider here only the total cross
section and study the finite width effects near threshold as well as
at very high energies. In fig.2 we show the total cross section in the
vicinity of the $W$ pair production threshold and the corrections
relative to the lowest order cross section due to finite width and QED
corrections in the region from 200~GeV to 1~TeV.  As it was pointed
out earlier the effects from the finite $W$ width do not vanish at high
energies but rather enhance the cross section by about 6\% at 1 TeV.

\begin{figure}[htb]
\begin{picture}(160,80)
\put(-5,0){\makebox(80,80){\epsfxsize=8cm \leavevmode
	\epsffile{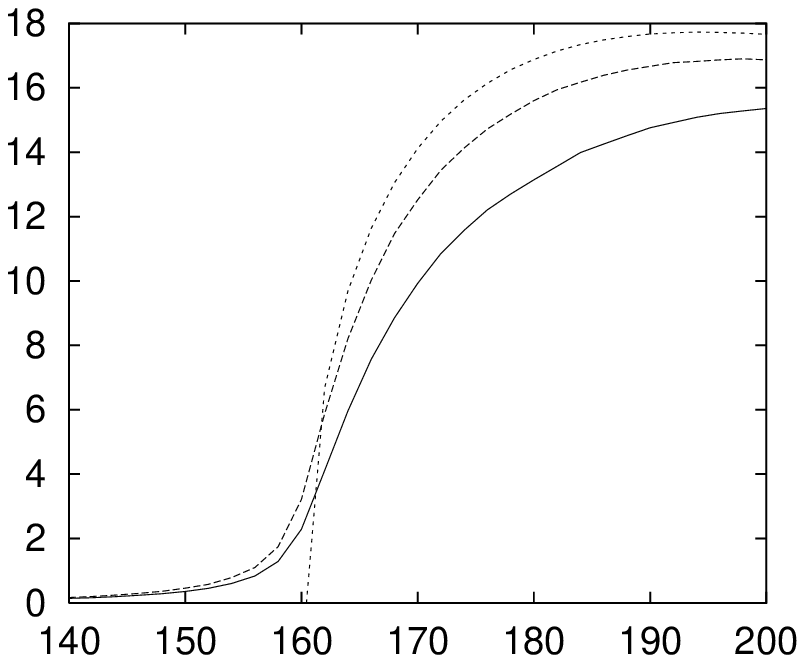}}}
\put(3,71){$\sigma$ [pb]}
\put(58,7){$\sqrt{s}$ [GeV]}
\put(75,0){
\makebox(80,80){\epsfxsize=8cm \leavevmode
	\epsffile{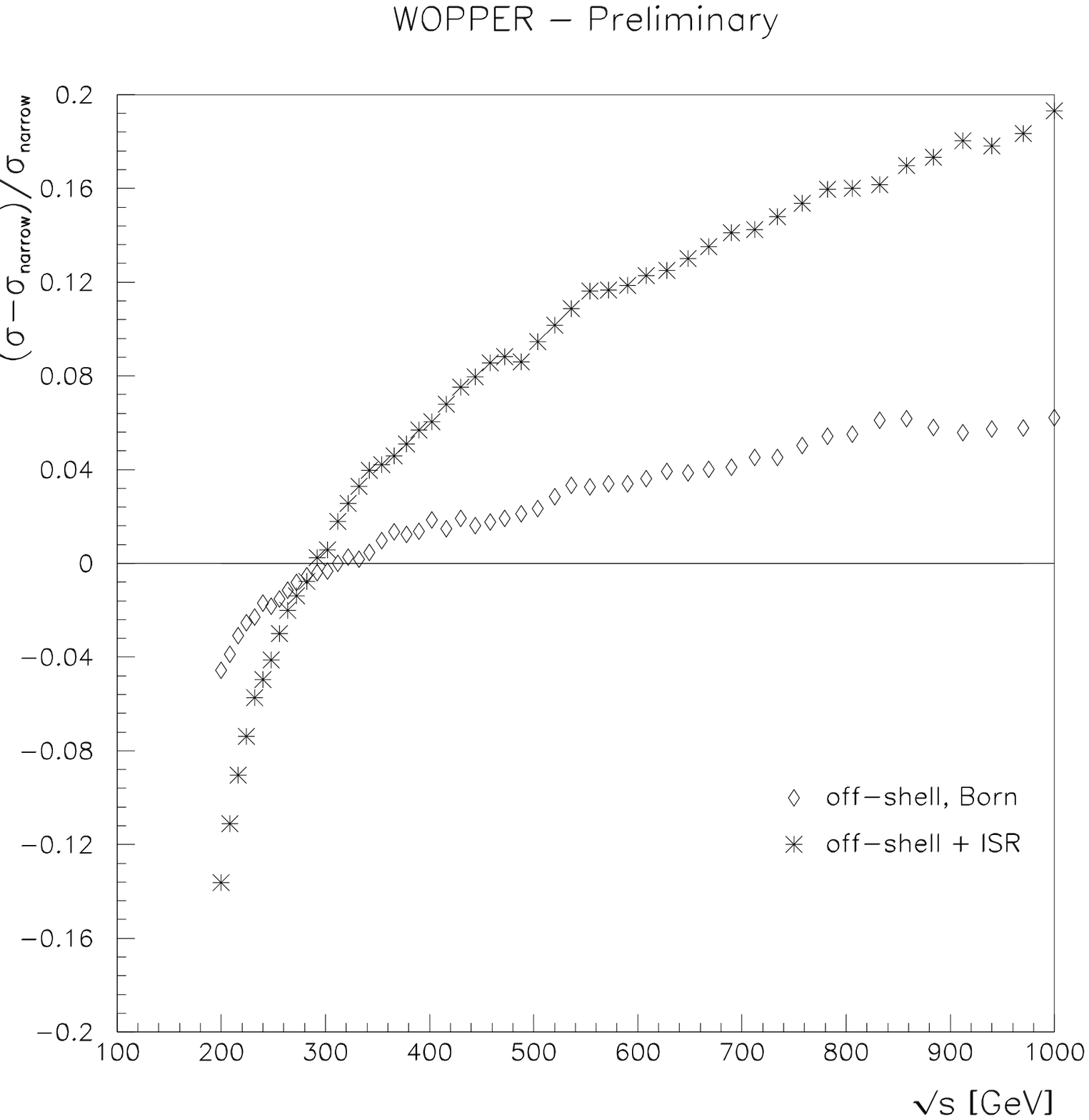}}}
\end{picture}
\caption{Left column: cross section in the threshold region (full
line: fully corrected, dashed line: finite width only, dotted line:
Born formula).
Right column: corrections relative to the Born cross section (stars:
fully corrected, open symbols: finite width only).}
\end{figure}



\begin{thebibliography}{99}

\bibitem{RC}
M.~B\"ohm, A.~Denner, T.~Sack, W.~Beenakker, F.~Berends, H.~Kuif,
{\em Nucl.\ Phys.\/} {\bf B304} (1988) 463;
J.~Fleischer, F.~Jegerlehner, M.~Zra\l{}ek,
{\em Z.~Phys.\/} {\bf C42} (1989) 409.

\bibitem{Finite-Width}
T.~Muta, R.~Najima, S.~Wakaizumi,
{\em Mod.\ Phys.\ Lett.\/} {\bf A1} (1986) 203;
D.~Bardin, M.~Bilenky, A.~Olchevski, T.~Riemann,
DESY~93-053, BI-TP~93/09, March 1993

\bibitem{CDMN91}
M.~Cacciari, A.~Deandrea, G.~Montagna, O.~Nicrosini,
{\em Z.~Phys.\/} {\bf C52} (1991) 421.

\bibitem{Structure}
E.A.~Kuraev, V.S.~Fadin,
{\em Yad.~Fiz.~\bf 41} (1985) 733;
G.~Altarelli, G.~Martinelli, in
J.~Ellis, R.~Peccei (eds.), {\em Physics at LEP\/},
CERN Report 86-02 (1986);
W.~Beenakker, F.A.~Berends, W.L.~van Neerven,
Contribution to the Ringberg workshop, April 1989.

\bibitem{QCD}
T.~Sj\"ostrand, Comp.~Phys.~Comp.~{\bf 39} (1986) 347.

\bibitem{Good-Stuff}
H. Anlauf et al., Darmstadt-Siegen Collaboration:
WOPPER -- A Monte Carlo event generator for
$W$ off-shell pair production including higher order QED corrections,
in preparation.

\bibitem{Width}
A. Aeppli, F. Cuypers, G.J. van Oldenborgh,
LMU-21/92, PSI-PR-93-05, TTP92-36

\bibitem{EE500}
Proc.\ of the Workshop -- Munich, Annecy, Hamburg --
Feb.~4 to Sep.~3, 1991, DESY 92-123

\end{thebibliography}
\end{document}